\documentclass[twocolumn,prb,twoside,a4paper]{revtex4}

\renewcommand{\familydefault}{\sfdefault}

\usepackage[font=footnotesize,labelfont=sf,bf,flushleft]{caption}
\usepackage{graphicx}
\usepackage[usenames,dvipsnames]{xcolor}
\usepackage{amsmath,bbm}

\usepackage[normalem]{ulem}
\usepackage{titlesec}
\titlespacing{\section}{0pt}{0pt}{-4pt}
\titlespacing{\subsection}{0pt}{0pt}{-4pt}

\newcommand{\white}[1]{{\color[rgb]{1,1,1}{#1}}}

\newcommand{\blue}[1]{{\color{black}{#1}}}

\renewcommand{\d}{\textrm{d}}

\titleformat{name=\section}{\filright \sf \normalsize \bfseries}{\thesection}{.5em}{}
\titleformat{name=\subsection}{\filright \sf \small \bfseries}{\thesubsection}{.5em}{}

\begin{document}
 

\renewcommand{\baselinestretch}{2}
\title{\flushleft{ \Huge{\bfseries Nanoscale temperature measurements using non-equilibrium Brownian dynamics of a levitated nanosphere\vspace{-0.4cm} }}}

\author{\hspace{-8.2cm} \bfseries J. Millen\footnotemark[2]\footnotemark[1], T. Deesuwan\footnotemark[2]\footnotemark[3], P. Barker\footnotemark[2] \& J. Anders\footnotemark[2]\footnotemark[4]\footnotemark[1]}

\affiliation{\white .  \vspace{-1.4cm} }

\maketitle

\renewcommand{\baselinestretch}{1}

\footnotetext[2]{Department of Physics and Astronomy, University College London, Gower Street, WC1E 6BT London, United Kingdom,}
\footnotetext[3]{Department of Physics, Imperial College London, Prince Consort Road, London SW7 2AZ, United Kingdom,}
\footnotetext[4]{Department of Physics and Astronomy, University of Exeter, Stocker Road, Exeter EX4 4QL, United Kingdom.}
\footnotetext[1]{j.millen@ucl.ac.uk, j.anders@exeter.ac.uk}

\small 

\noindent {\bfseries
Einstein realised that the fluctuations of a Brownian particle can be used to ascertain properties of its environment\cite{Einstein05}. A large number of experiments have since exploited the Brownian motion of colloidal particles for studies of dissipative processes\cite{pinknoisepaper, Berut2012}, providing insight into soft matter physics\cite{GB2009,ADY87, Florin2001}, and leading to applications from energy harvesting to medical imaging\cite{energyharvesting, medicalimaging}. Here we use optically levitated nanospheres that are heated to investigate the non-equilibrium properties of the gas surrounding them. Analysing the sphere's Brownian motion allows us to determine the temperature of the centre-of-mass motion of the sphere, its surface temperature and the heated gas temperature in two spatial dimensions. We observe asymmetric heating of the sphere and gas, with temperatures reaching the melting point of the material. This method offers new opportunities for accurate temperature measurements with spatial resolution on the nanoscale, and a new means for testing non-equilibrium thermodynamics.}

\rm

Brownian motion is an essential tool for the investigation of micro and nanoscale properties and processes in the physical and life sciences. Experiments are now at the nanoscale, where single particles can be strongly trapped by optical tweezers, electrostatic and magnetic traps{\cite{Li10,HK91}. The precise control offered by these techniques has made possible the experimental observation of physical phenomena at previously inaccessible scales. These include the measurement of the non-Gaussian nature of fluids \cite{pinknoisepaper} and the verification of Landauer's principle\cite{Berut2012}, which links thermodynamics with information theory. Trapping methods have also allowed the observation of bio-molecular forces\cite{ADY87,GB2009}, the accurate determination of nanoscale particle properties, such as charge and size\cite{Mojarad12}, and the characterisation of  potential landscapes when modified by, for example, colloidal interactions\cite{Florin2001}. These experiments have focussed on the particle's motion and its interaction with the bath; however, the substantial impact of a trapped particle's surface temperature on its dynamics has not been explored. Observing Brownian dynamics at the nanoscale when large temperature differences are present is difficult and has only recently been achieved\cite{Rings2010, TwoConductorsAtdifferentTemps}. Here we report on the measurement of the modified Brownian motion of a strongly heated trapped nanosphere in the underdamped (Knudsen) regime, which produces a non-equilibrium gas surrounding it. A schematic of the experiment is shown in Fig.~\ref{fig:setup}a together with a photograph of the trap in Fig.~\ref{fig:setup}b.

\begin{figure}[t]
	 {\includegraphics[width=0.46\textwidth]{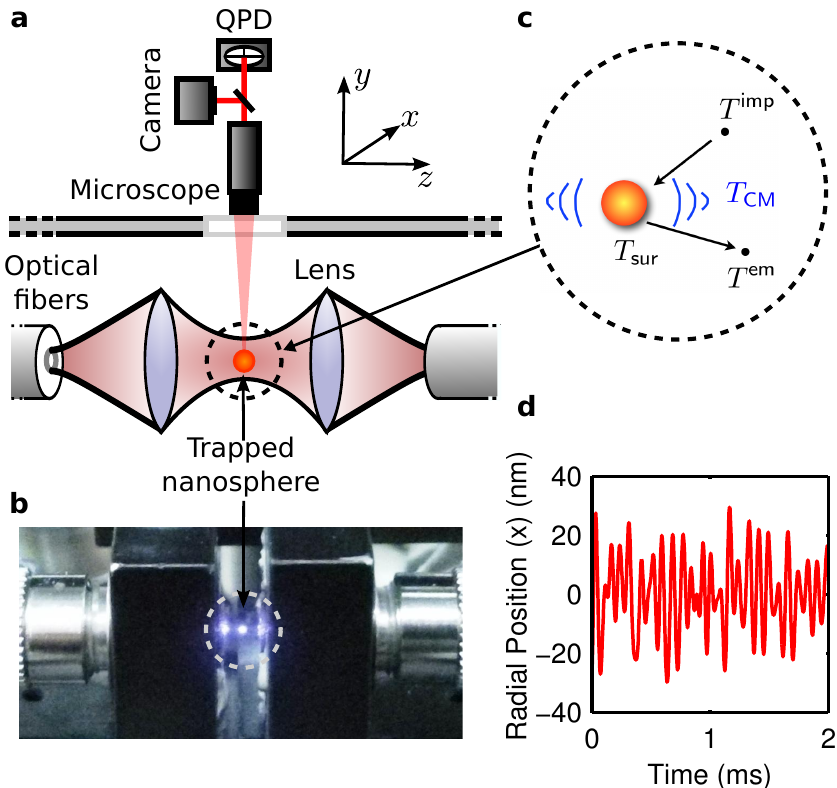}}	
\caption{\label{fig:setup} 
\footnotesize {\bfseries Experimental schematic and photograph of levitation experiment.} 
{\bfseries a,} Silica spheres are levitated in a dual-beam optical tweezer inside a vacuum chamber. Light of wavelength 1064~nm is coupled into lenses from single-mode optical fibers, creating an optical trap. The motion of the levitated sphere is monitored with a camera and a quadrant photodetector (QPD). 
{\bfseries b,} Photograph of the trapping region with trapped sphere visibly scattering trap light.
{\bfseries c,} Schematic showing the temperatures involved in a collision with a heated sphere: the sphere's centre-of-mass temperature ($T_{\sf CM}$) and surface temperature ($T_{\sf sur}$), and the temperatures of the impinging gas particles ($T^{\sf imp}$) and emerging gas particles ($T^{\sf em} $) with $T^{\sf imp} < T_{\sf CM}< T^{\sf em} < T_{\sf sur}$. 
{\bfseries d,} Typical Brownian motion position trace for a 105.1~nm sphere at 1~mbar. 
}
\end{figure}

\begin{figure*}[t]
	{\includegraphics[width=0.98\textwidth]{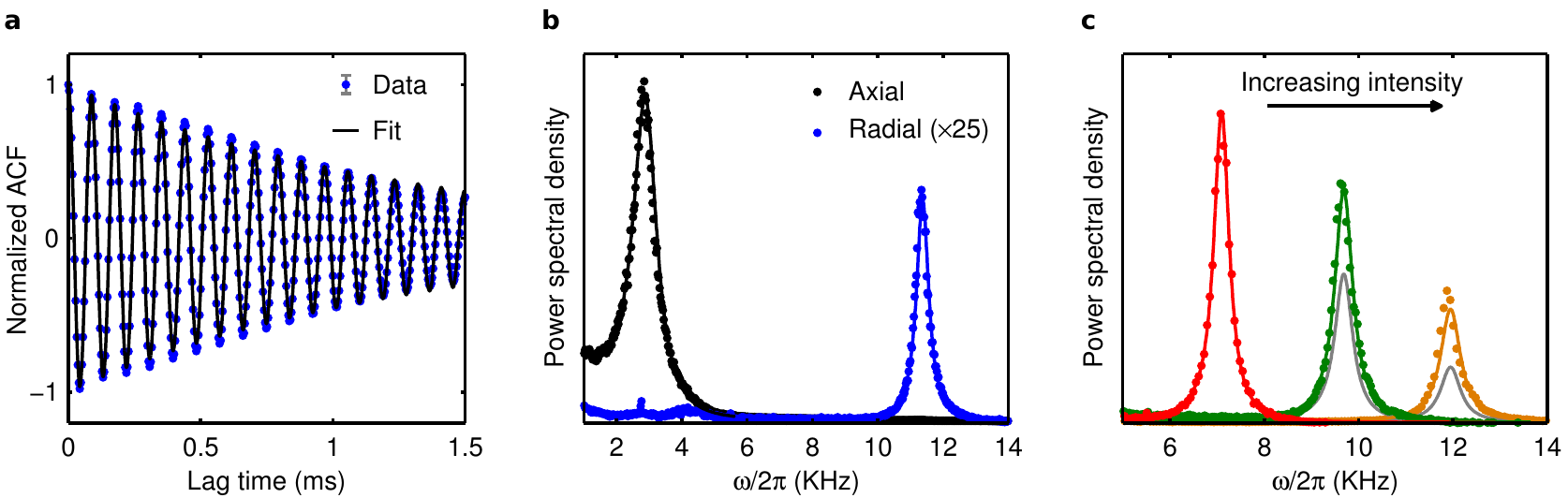}}
\caption{\label{fig:powerspec} 
\footnotesize {\bfseries Two dimensional Brownian motion and its variation with trap laser intensity.} All displayed data are for 105.1~nm spheres at 1~mbar.
{\bfseries a,}  Autocorrelation function (ACF) of experimentally measured radial position  (blue points with grey error bars) and fit with theory (black solid line) (\blue{supplementary information equation S4}).
{\bfseries b,} Measured axial (black points) and radial position (blue points) power spectra with fits to $P(\omega)$ (solid lines) over trap frequency ${\omega \over 2 \pi}$. Because of the low damping at mbar pressures, the axial ($z$) and radial ($x$) peak frequencies, $\tilde{\omega}_{x, z} = \sqrt{\omega_{x, z}^2 - {\Gamma_{\sf CM}^2/2}}$, approximate the trap frequencies $\omega_{x, z}$.
{\bfseries c,} Measured radial power spectra (points) with fits (solid lines) for three values of trapping laser intensity. From left to right the intensity of the trapping laser beams is increased by a factor of 2.1 and the peak frequency ${\tilde{\omega}_x/2 \pi}$ increases with intensity, as expected. If there were no temperature increase from the initial intensity (red) the peak height would behave as indicated by the grey solid lines. However, it can be seen that the height of the peak, $A/\Gamma_{\sf CM} \omega_x^2 = 2 k_B T_{\sf CM}/ M \Gamma_{\sf CM} \omega_x^2$, increases with intensity in comparison to the grey lines, indicating the increase of the centre-of-mass temperature, $T_{\sf CM}$, of the sphere.}
\end{figure*}

When the temperature of the surface of a sphere is different from its surrounding gas, for example when it is heated by absorption of light, heat is transferred to the colliding gas particles. Impinging gas particles at temperature $T^{\sf imp}$ do not equilibrate to the same temperature as the sphere surface at temperature $T_{\sf sur}$ (partial accommodation\cite{Goodman1980}). They emerge with different energy, but can be assumed to be thermally distributed for highly inelastic collisions\cite{Epstein24,BCF90} with a different temperature, $T^{\sf em} \not = T^{\sf imp}$ (Fig.~\ref{fig:setup}c). The change in emerging temperature implies that the gas is not in  thermal equilibrium, as assumed in the analysis of most optical tweezer experiments. In contrast, we derive a model of modified Brownian motion based on two non-interacting baths, the impinging gas  at temperature $T^{\sf imp}$ and the emerging gas at $T^{\sf em}$, which interact with the sphere's centre-of-mass motion (see \blue{supplementary information}). This situation only occurs in the Knudsen regime, where the mean free path of the gas is much larger than the sphere radius and the gases do not equilibrate in the vicinity of the sphere.
The sphere's centre-of-mass motion adopts a non-equilibrium steady state that mediates heat transfer between the two baths. Importantly, even though the gas is not in equilibrium the power spectrum of the sphere's fluctuating position is of thermal form, $ P(\omega) = {2 k_B T_{\sf CM} \over M} \, {\Gamma_{\sf CM} \over (\omega_0^2 - \omega^2)^2 + \omega^2 \Gamma_{\sf CM}^2}$, where $M$ is the mass of the sphere and $\omega_0$ its trap frequency. The power spectrum is parametrised by an \emph{effective} centre-of-mass temperature, $T_{\sf CM} :=  {T^{\sf imp} \Gamma^{\sf imp} + T^{\sf em} \Gamma^{\sf em} \over \Gamma^{\sf em} + \Gamma^{\sf imp}}$, and an \emph{effective} damping rate, $\Gamma_{\sf CM} := \Gamma^{\sf em} + \Gamma^{\sf imp}$ (see \blue{supplementary information}).

To study the Brownian fluctuations in thermal non-equilibrium we use silica spheres of 105.1~nm and 2.56~$\mu$m diameter in a dual-beam optical trap with wavelength 1064~nm. Silica can be heated due to the absorption of trap light by impurities in the material. At any given pressure, the temperature of the spheres can be controlled through the laser light intensity with an approximately linear dependence\cite{Peterman03}, $T_{\sf sur} \propto I $. In the steady state, $T_{\sf sur}$ is given by the balance between heating due to laser absorption, and cooling due to collisions with the gas and the emission of blackbody radiation. The exact rate of heating depends on the prevalence of impurities within the material and thus varies from sphere to sphere. Even though absorption is low, high temperatures ($\approx$2000~K) can be reached because of the poor heat transfer to the surrounding gas at low pressures. 

The optical trap is well approximated by a three-dimensional harmonic potential with frequencies in the direction along the laser focus, $\omega_z$, and those orthogonal to it, $\omega_x$ and $\omega_y$. Frequencies range from approximately 1~kHz to 15~kHz and depend on laser intensity, the axis of motion, and particle size. The trap is placed inside a vacuum chamber and the motion of the particle is monitored \blue{(see Methods)} in two spatial directions, axial ($z$) and radial ($x$), using a quadrant photo-detector (Fig.~\ref{fig:setup}a). The particles are stable in the trap for many hours.

We vary the laser intensity from 0.3-15~MW cm$^{-2}$ and lower the pressure of the gas in the chamber from atmospheric to approximately 1~mbar. The radial position of a 105.1~nm diameter sphere as a function of time at 1~mbar is depicted in Fig.\ref{fig:setup}d, showing the harmonic motion with Brownian fluctuations. From the axial and radial position data we calculate and fit the autocorrelation function (Fig.~\ref{fig:powerspec}a) and the power spectra (Fig.~\ref{fig:powerspec}b). Fitting the experimental power spectra to $P(\omega)$ allows the determination of the effective damping coefficient, $\Gamma_{\sf CM}/(2 \pi)$, and the prefactor, $A = {2 k_B T_{\sf CM} \over M}$, for each spatial direction once a voltage to position calibration factor has been determined for the QPD \blue{(see Methods)}. With this method the ratio of the effective temperature, $T_{\sf CM}$, over mass, $M$, can be determined for any laser intensity and pressure. As expected, the trap frequency increases with increasing laser intensity (Fig.~\ref{fig:powerspec}c). In addition the sphere's centre-of-mass motion experiences a substantial increase in temperature, as can be seen in Fig.~\ref{fig:powerspec}c, when compared to a power spectra at increased frequency but remaining at the initial temperature.

\begin{figure*}[t]
	{\includegraphics[width=0.98 \textwidth]{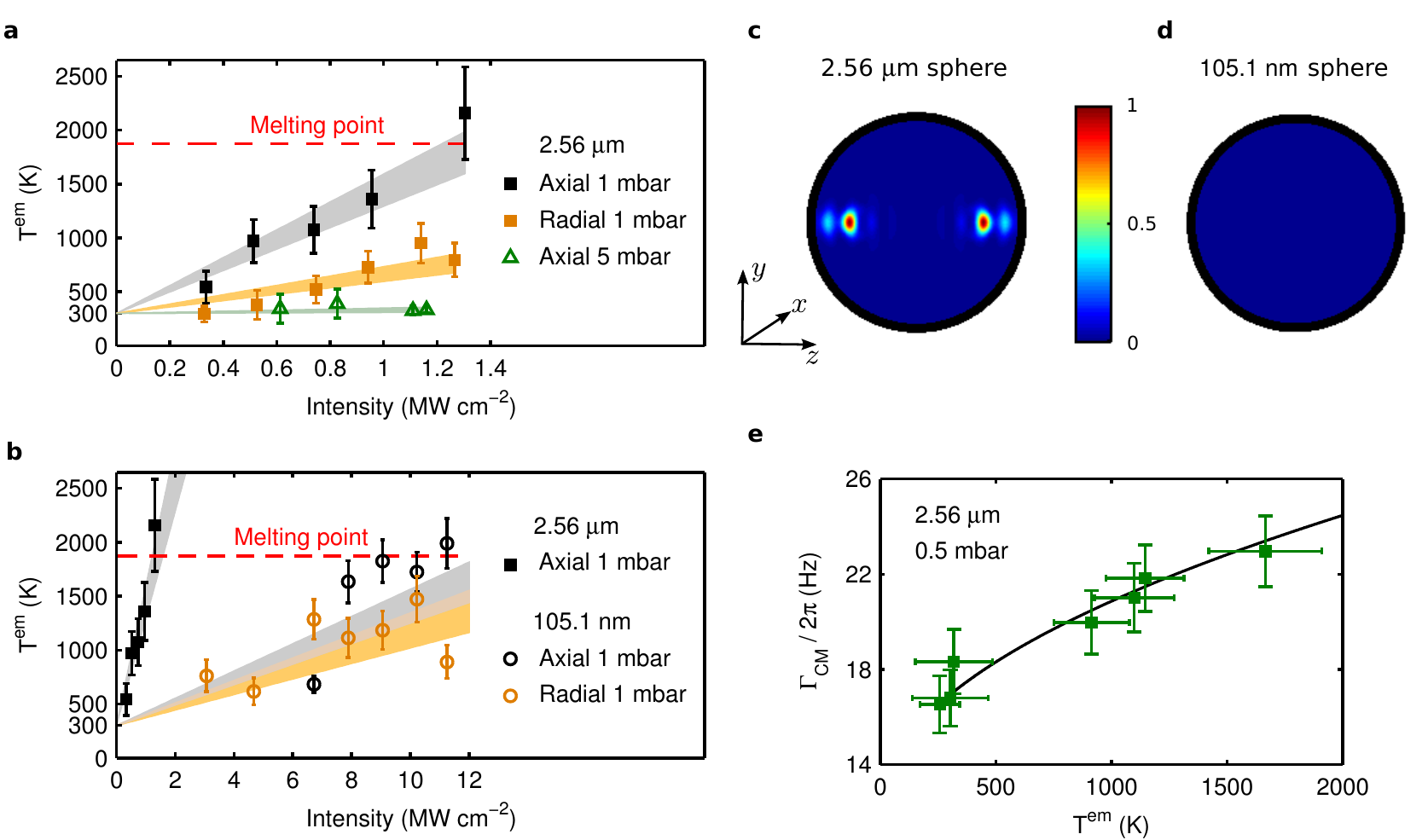}} 
\caption{ \label{fig:heat} 
\footnotesize {\bfseries Analysis of heating as a function of pressure, spatial direction and sphere size.} 
{\bfseries a,} Temperature of emerging gas particles with trapping laser intensity for the large spheres in the axial (black squares) and radial directions (orange squares) at 1~mbar, and in the axial direction at 5~mbar (green triangles). Asymmetric heating is observed at 1~mbar while no heating occurs at 5~mbar. The shaded areas show linear fits with one standard deviation of uncertainty indicating that a linear relationship $T^{\sf em} \propto T_{\sf sur} \propto I$ holds, as expected\cite{Peterman03}.
{\bfseries b,} Temperature of emerging gas particles ${T}^{\sf em}$ with trapping laser intensity, for 105.1~nm diameter (open circles) and 2.56~$\mu$m diameter spheres (filled squares) at 1~mbar in the axial (black) and radial (orange) direction. Heating of the gas is clearly observed for both large and small spheres. The shaded areas show linear fits with  one standard deviation of uncertainty as a guide to the eye.
{\bfseries c,} Field intensity inside the 2.56~$\mu$m spheres for two incident counterpropagating Gaussian beams clearly showing lensing of the light. 
{\bfseries d,} Field intensity inside the 105.1~nm spheres for two incident counterpropagating Gaussian beams, showing a homogenous intensity throughout the spheres. 
{\bfseries e,} Measured (green squares) variation of the damping coefficient $\Gamma_{\sf CM}$ for a 2.56~$\mu$m diameter sphere at 0.5~mbar over temperature of the emerging gas particles, $T^{\sf em}$. Also shown is the theoretical $\Gamma_{\sf CM}$ (black solid line) with no free parameters, using the experimentally given parameters $T^{\sf imp} = $ 294~K and $\Gamma^{\sf imp} =$ 76.3~Hz for 2.56~$\mu$m spheres.
}
\end{figure*}

With the measured centre-of-mass temperature, $T_{\sf CM}$,  the temperature of the emerging gas particles, $T^{\sf em}$,  can be deduced using the fact that the gas particles impinge at room temperature, $T^{\sf imp} = $ 294~K. Also needed to extract $T^{\sf em}$ is the damping rate of the sphere due to the emerging gas which we have derived in the \blue{supplementary information}. It is given by $\Gamma^{\sf em}=  {\pi \over 8} \sqrt{T^{\sf em} \over T^{\sf im}} \, \Gamma^{\sf imp}$, where $\Gamma^{\sf imp} = {4 \pi \over 3} {m N R^2 \bar{v}_{T^{\sf imp}} \over M}$ with $m$ the molecular mass, $\bar{v}_{T^{\sf imp}} = \sqrt{8 k_B T^{\sf imp} \over \pi m }$ the mean thermal velocity, $N$ the gas particle density and $R$ the sphere radius\cite{Epstein24}. At the millibar pressures used in the experiment corrections due to small Knudsen numbers\cite{BCF90} can be neglected. The temperature of the emerging gas particles can now be found by solving $T_{\sf CM} = {(T^{\sf imp})^{3/2}  + {\pi \over 8} (T^{\sf em})^{3/2}  \over (T^{\sf imp})^{1/2} + {\pi \over 8} (T^{\sf em})^{1/2} }$ for  $T^{\sf em}$. 

Figures~\ref{fig:heat}a,b show the inferred temperature of the emerging gas surrounding a 2.56~$\mu$m and a 105.1~nm diameter sphere as a function of laser intensity. The typical accuracy of the temperature measurement is 10\%. For the 2.56~$\mu$m sphere (Fig.~\ref{fig:heat}a) strong gas heating is always observed at 1~mbar while no measurable heating occurs at 5~mbar. In the latter case, the more frequent collisions of the sphere with gas particles take heat from the sphere surface away more efficiently. This prevents heating of the sphere surface and the gas emerging from it. Since the larger spheres act as a lens for the trapping light a non-uniform intensity profile is expected within it. This is illustrated by our simulations, shown in Fig.~\ref{fig:heat}c, using generalised Lorentz-Mie theory and the \emph{Optical Tweezers Toolbox}\cite{Optics}. For particles of this size one therefore expects an asymmetric surface temperature which would lead to a higher on-axis temperature of the emerging gas particles. Indeed, we measure a considerably larger temperature in the axial direction than the radial direction (Fig.~\ref{fig:heat}a) giving unambiguous evidence that  at 1~mbar the large sphere is moving through two baths, the impinging gas at room temperature and the emerging gas with spatially varying temperature. 

The gas surrounding the smaller spheres generally experiences a much slower heating when compared to the larger spheres (Fig.~\ref{fig:heat}b). This can be understood by realising that while cooling mechanisms such as collisions and blackbody radiation scale with surface area, heating due to absorption from the laser beam scales with volume. The smaller spheres thus have an area to volume ratio that leads to much more efficient cooling than the larger spheres. Heating depends on the number of impurities in the material and the smaller spheres show a much more varied heating pattern than the larger spheres. The data displayed in Fig.~\ref{fig:heat}b shows one of the strongest heating observed for the small spheres. In contrast to the larger spheres, no lensing effects should occur for the smaller 105.1~nm spheres as shown by our simulation (Fig.~\ref{fig:heat}d). This agrees well with the experimental data which do not indicate significant spatial asymmetry (Fig.~\ref{fig:heat}b).

Figure~\ref{fig:heat}e shows the variation of the damping coefficient $\Gamma_{\sf CM}$ with emerging gas temperature, $T^{\sf em}$, together with the theory result for $\Gamma_{\sf CM}$  without any fitting parameters, showing the consistency of the theoretical prediction. Due to $\Gamma^{\sf em}$'s moderate temperature dependence on $T^{\sf em}$ a large increase in temperature is required before any significant change in the damping is seen. The relationship between the gas temperatures and the surface temperature of the sphere is given by a material constant, the accommodation coefficient $\alpha= {T^{\sf em} - T^{\sf imp} \over T_{\sf sur} - T^{\sf imp}}$. The accommodation coefficient for silica is known\cite{Ganat11} and close to 0.777 for moderate surface temperatures around 300~K. This allows us to infer the surface temperature of the nanospheres, $T_{\sf sur} = 294~K + {T^{\sf em} - 294~K \over 0.777}$. At high temperatures the accommodation coefficient is not known and the emerging gas temperature establishes a lower bound on the surface temperature. However, independent measurements of the surface temperature (based on the emitted blackbody radiation) are possible in this regime\cite{Landstrom04}, which would allow the determination of the variation of the accommodation coefficient.

\begin{figure}[t]
	{\includegraphics[width=0.47 \textwidth]{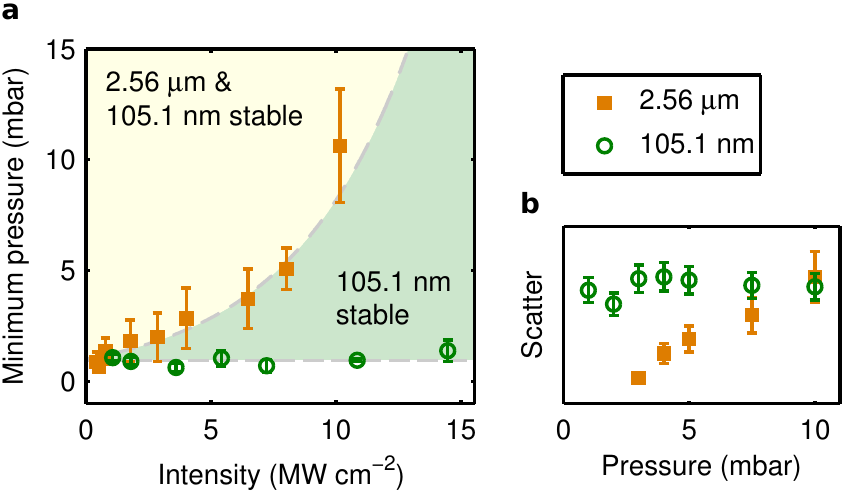}}
\caption{\label{fig:melting}
\footnotesize
{\bfseries Minimum attainable pressure.} 
{\bfseries a,} Minimum pressure for which 2.56~$\mu$m spheres (orange squares) and 105.1~nm spheres (green  circles) were stably trapped, versus intensity. This data is averaged over different spheres of each size ($>$5) at each intensity. Due to their strong heating the 2.56~$\mu$m spheres evaporate at high intensities, leading to an intensity dependent attainable pressure. In contrast, the smaller spheres are lost below an average pressure of  0.95~mbar regardless of the laser intensity. 
{\bfseries b,} Scatter plot over pressure at fixed intensity 1.3~MW cm$^{-2}$ showing the total scattered light intensity from the 2.56~$\mu$m spheres (orange squares) and 105.1~nm spheres (green circles). At lower pressure, the scatter drops strongly for the 2.56~$\mu$m spheres indicating evaporation, while the 105.1~nm spheres show an approximately constant scatter.}
\end{figure}

Finally, experiments old \cite{Ashkin76} and new \cite{Burnham10,Monteiro13} with levitated particles have faced difficulties reaching low pressures, and it has been widely believed that heating plays a part in this \cite{Ashkin76,Burnham10}. Our data imply that heating is the dominant cause for the loss of larger particles from the trap. However, temperature independent mechanisms also play an important role. As indicated in Fig.~\ref{fig:heat}a for the 2.56~$\mu$m spheres the emerging gas temperature approaches the melting point of silica. This means that the spheres can undergo phase changes at high enough intensities, eventually boiling and vanishing from the trap. Such loss would be evidenced by a marked rise of the minimum pressure at which the spheres can still be trapped for increasing intensity. For a higher laser intensity, where there is more heating, a faster cooling rate (higher pressure) is required to balance the surface temperature of the sphere. The measured minimum attainable pressure with trapping laser intensity is displayed in Fig.~\ref{fig:melting}a and shows exactly this behaviour for the 2.56~$\mu$m spheres. In addition we observe that at fixed laser intensity the level of light scattered from the spheres decays strongly when the pressure is reduced (Fig.~\ref{fig:melting}b) indicating that evaporation occurs. In contrast, Fig.~\ref{fig:melting}a shows that the loss of the smaller spheres at low pressures in the present experiment is due to a temperature-independent mechanism. For example parametric heating that results from any noise in the system, such as laser noise, vibrations etc., can render the trap unstable. Consistent with this interpretation, it was  observed\cite{Monteiro13}  that at low trap laser intensities the limiting pressure is independent of size for a range of sphere diameters from 50\,nm to 2.56\,$\mu$m. It is known that feedback cooling is a reliable method well suited to correct for parametric heating \cite{Li11, Gieseler13}. 

We have derived and tested a model for the Brownian motion of a particle in a non-equilibrium gas in the Knudsen regime. Using an optically levitated nanosphere we have performed measurements of the temperature of the non-equilibrium gas surrounding the spheres at mbar pressures with a typical precision of 10\%. The experiment allowed us to observe substantial heating of the gas particles emerging from the sphere surface even for modest trapping laser intensities. Our method can also be used with other trapping techniques, such as as electrostatic or magnetic levitation\cite{HK91}, where the particle is heated by other means. An anisotropic temperature distribution of the gas surrounding the larger spheres was found, giving testimony to the non-equilibrium nature of the sphere's centre-of-mass motion. We achieve a spatial resolution of a few nm which is greater than standard optical imaging techniques. While the fluctuations of the sphere were here used to provide information about the bath, they are also a means to determine properties of the trapped sphere, such as its surface temperature and its accommodation coefficient. Our results show that a careful distinction between the operating temperature, e.g. room temperature, the spatially varying surface temperature and the sphere's three centre-of-mass temperatures must be considered in any optical tweezer experiment. This is particularly relevant for recent experiments and proposals\cite{Li11, Gieseler13, Chang10, Romero10} to cool the centre-of-mass motion of optically levitated particles with the aim of reaching the quantum ground state. Using nanospheres as a local probe of bath properties in other non-equilibrium environments in the Knudsen regime is a promising tool for micro-rheology\cite{Bennett2013} and micro-fluidics\cite{microfluidics}, and for the study of aerosols\cite{Burnham10}. For instance, with nanospheres of smaller size, $\approx$ 10~nm, this method could be employed at nearly atmospheric pressures. Finally, the presented approach provides a starting point for investigations of non-equilibrium thermodynamics extending into the underdamped regime, such as testing fluctuation theorems, studying stochastic thermodynamics\cite{Seifert2008} and entropy production\cite{TwoConductorsAtdifferentTemps}, and exploring the link between thermodynamics and information theory\cite{Berut2012}.

\setlength{\parskip}{1ex}
\footnotesize 
\renewcommand{\familydefault}{\rmdefault}

\section*{Methods} 

\noindent {\bfseries Trap setup and characterisation.} 
Silica spheres of diameter 105.1~nm (Corpuscular Inc.) and 2.56~$\mu$m (Bangs Labs) are trapped. The trap light is split into two beams of equal power by an acousto-optic modulator (AOM), introducing a frequency difference of 80~MHz and eliminating interference effects in the trap. These two beams are coupled into separate single mode optical fibers, which enter a vacuum chamber, with the output focussed by aspheric lenses of focal length 1.45~mm to create a trap with a beam waist of approximately 1~$\mu$m. One of the fiber-lens systems is mounted on a translation stage allowing alignment in three dimensions. The spheres are introduced to the trapping region using a nebulizer (Omron NEU22) at atmospheric pressure, as the damping force of the air is required to load the trap. The spheres are suspended in methanol with very low particle concentration, and the solution is sonicated in an ultrasonic bath for at least an hour before trapping to prevent clumping. 

\noindent {\bfseries Monitoring.} 
The trapped particle is monitored from outside the vacuum chamber by a CMOS camera (for diagnostics) and a quadrant photodetector (QPD) using a microscope with a zoom of 45$\times$ (Fig.~\ref{fig:setup}a). The detectors pick up scattered trapping light. When very low trap light intensities are required (for the 2.56~$\mu$m spheres), a weak beam of 532~nm is used to illuminate the sphere, with an intensity of $\sim 0.5~$kW\,cm$^{-2}$, as the detector is more sensitive at this wavelength. The QPD allows us to monitor the position fluctuations of the trapped sphere in the axial ($z$) and radial ($x$) direction. The orthogonal direction ($y$) offers the same trapping landscape as the radial direction when the system is well aligned. Since the QPD measures intensity fluctuations on its pixels as the particle moves, we are able to measure movement with nanometer resolution.

\noindent {\bfseries Data acquisition and analysis.}
The pressure in the trap is measured at 15\% accuracy with a Pirani and Bourdon gauge. The trapping light intensity is varied with an AOM, and also monitored by collecting light which is scattered by the sphere. An oscilloscope records the motion of each trapped sphere at a fixed pressure for 20 seconds, with a time resolution of $4~\mu$s. The power spectral density $P(\omega)$ of the motional data is calculated, and analyzed using the equation in the text, with an additional constant valued fitting parameter to account for a non-zero background. 
To calibrate the QPD data sets are taken at high enough pressures such that no variation in $T_{\sf CM}$ is seen with laser intensity, indicating that the gas surrounding the sphere is in thermal equilibrium at room temperature, 294~K. Each nanosphere behaves in a different way, due to variation in manufacture. This means that for a given intensity and pressure, different spheres of the same size will reach different temperatures. However, the trends are consistent across all realisations.

\noindent {\bfseries Theoretical model.} The sphere is assumed to interact in each spatial direction, $x$, with two independent  baths, the impinging gas and the emerging gas. Collisions between gas particles and sphere are inelastic, i.e. the sphere has a large accommodation coefficient, the particles stick on the surface and emerge independent in time and energy from the impinging gas particles. Both the impinging and emerging gas are assumed thermally distributed\cite{Epstein24, BCF90} with temperatures $T^{\sf imp}$ and $T^{\sf em}$, respectively. The sphere experiences an effective damping force against its motion with damping coefficient  $\Gamma_{\sf CM} = \Gamma^{\sf imp} +\Gamma_x^{\sf em}$, with  the known\cite{Epstein24} impinging damping coefficient, $\Gamma^{\sf imp}$. For the emerging gas damping coefficient we consider the total drag force due to each leaving particle on the sphere moving with velocity $V$ in three dimensions, $F^{\sf em}_{\sf drag} =  \int_{0}^{\pi} \int_{0}^{2 \pi} \, {m N \sqrt{\pi} \, \cos \theta \, V \over 4 \,  \sqrt{ m \over 2 k_B T^{\sf em}} } \,  \cos \theta \, \d S$, from which we obtain $\Gamma^{\sf em} = {F^{\sf em}_{\sf drag} \over M V}$. Details of the derivation are given in the supplementary information.


\section*{Acknowledgements}  
\noindent We thank Ian Ford (IF) for insightful discussions, and Isabel Llorente Garcia, Dorothy Duffy and IF for critical reading of the manuscript. J.M. and P.B. acknowledge funding from EPSRC EP/H050434/1. T.D. is supported by the Royal Thai Government and the EPSRC. J.A. acknowledges funding from the Royal Society. This work was supported by the European COST network MP1209. 

\section*{Author contributions} 
\noindent J.M. and P.B. designed the experiments. J.M. performed the experiments, analysed the data and performed the error analysis. T.D and J.A. developed the two bath model. J.A. derived the damping rate. P.B. performed the field simulation. All authors contributed to the data analysis and wrote the manuscript.

\section*{Competing financial interests} 
\noindent The authors declare no competing financial interests. 

\end{document}